\documentclass[11pt]{article}
\usepackage{amsmath}
\usepackage{amsfonts}
\usepackage{amssymb}
\begin{document}
\title{\textbf{Reconstruction Procedure for Nonlocal Gauss--Bonnet Models}}
\author{Emilio~Elizalde$^{1}$\footnote{E-mail address:
elizalde@ieec.uab.es}, \
Ekaterina~O.~Pozdeeva$^{2}$\footnote{E-mail address:
pozdeeva@www-hep.sinp.msu.ru}, \
Sergey~Yu.~Vernov$^{2}$\footnote{E-mail address:
svernov@theory.sinp.msu.ru}\vspace*{3mm} \\
\small $^1$Instituto de Ciencias del Espacio (ICE/CSIC) \, and
\\
\small  Institut d'Estudis Espacials de Catalunya (IEEC) \\
\small Campus UAB, Carrer de Can Magrans, s/n \\
\small  08193, Bellaterra (Barcelona), Spain\\
\small  $^3$Skobeltsyn Institute of Nuclear Physics, Lomonosov
Moscow State University,\\
\small  Leninskie Gory, GSP-1, 119991, Moscow, Russia}

\date{ \ }
\maketitle

\begin{abstract}
We investigate the cosmological dynamics of
nonlocally corrected gravity involving a function of the inverse
d'Alembertian acting on the Gauss--Bonnet term. Casting the dynamical equations in local form, we derive the reconstruction procedure. We find  conditions on the model parameters  that are sufficient for the existence of de Sitter solutions and obtain these solutions explicitly.
\end{abstract}


\section{Nonlocal models with the Gauss--Bonnet term}

A model that includes in its action a nonlocal term $R f(\Box^{-1}R)$, where $f$ is a function of the inverse d'Alembertian  $\Box^{-1}$,  has been proposed in Ref.~\cite{Deser:2007jk} to explain the current cosmic acceleration without dark energy\cite{Jhingan:2008ym,Woodard,Deser:2013uya} (see also Ref.~\cite{Woodard:2014iga,Maggiore:2016gpx,Deser:2019lmm}). This model has a local formulation~\cite{Odintsov0708} as a model with additional scalar fields nonminimally coupled to gravity. In Refs.~\cite{Deser:2013uya,Zhang:2016ykx,Park:2017zls} the relation between the original nonlocal equations and their local representation, and the differences between these formulations, have been actively discussed. For the local version, the perturbation analysis has been proposed\cite{Koivisto,Koivisto2}.
In Refs.~\cite{Koivisto,EPV1,SYuV,EPV2}, reconstruction procedures for the local representation have been discussed that allow to find such a function $f$ so that the model has interesting cosmological solutions, for example,  de Sitter and power-law ones~\cite{EPV1,SYuV,EPV2,EPV3}, in the analytic form.
The local formulation of this model with an additional Gauss--Bonnet term has been explored in Ref.~\cite{Non-local-FR}.

In Ref.~\cite{Capozziello:2008gu}, nonlocal models with the Gauss--Bonnet term were proposed in its general form and the localization procedure, which transforms these nonlocal models into models of the Einstein--Gauss--Bonnet gravity with scalar fields, were considered.

In Refs.~\cite{Cognola:2009jx,Koshelev:2013lfm,Elizalde:2018qbm}, the investigation of nonlocal models with the Gauss--Bonnet term have been continued, and new analytic solutions have been found.

In this paper, we consider the following type of nonlocal models with the Gauss--Bonnet term $\mathcal{G}$:
\begin{equation}
S_{NL}=\int d^4x\sqrt{-g}\left[\frac{M^2_{\mathrm{Pl}}}{2}R+\mathcal{G}f\left(\frac{1}{\mathcal{M}^2}\Box^{-1}
\mathcal{G}\right)-\Lambda+\mathcal{L}_m\right]\,,
\end{equation}
where $M_{\mathrm{Pl}}$ is the reduced Planck mass, $f$ a differential function, $\mathcal{M}$ and $\Lambda$ are constants, and the Gauss--Bonnet term is given by
$$\mathcal{G}=R^2-4R_{\mu\nu}R^{\mu\nu}+R_{\mu\nu\alpha\beta}R^{\mu\nu\alpha\beta}\,.$$

A localization procedure~\cite{Odintsov0708,Capozziello:2008gu} allows us to get a local model with the Gauss--Bonnet term, by introducing scalar fields
$\xi$ and $\psi$, and rewriting the action $S_{NL}$ as
\begin{equation}
S_{L}=\int d^4x\sqrt{-g}\left(\frac{M^2_{\mathrm{Pl}}}{2}R+f(\psi) \mathcal{G}+\xi\left(\mathcal{G}-\mathcal{M}^2\Box\psi\right)-\Lambda+\mathcal{L}_m\right).
\end{equation}
Varying $S_L$ over $\xi$, we get
\begin{equation}
\label{equpsi}
 \Box\psi=\frac{\mathcal{G}}{\mathcal{M}^2}\quad\Rightarrow\quad \psi=\frac{\Box^{-1}\mathcal{G}}{\mathcal{M}^2}\,,
\end{equation}
what allows to reconstruct the initial nonlocal action.

To consider cosmological solutions, we assume a spatially
flat Friedmann--Lema\^{i}tre--Robertson--Walker metric, with
$$ds^2={}-dt^2+a^2(t)\delta_{ij}dx^idx^j\,,$$
and consider scalar fields that depend on time only. Using  ${\cal G}=24H^2(\dot H +H^2)$, with the Hubble parameter $H=\dot{a}/a$,
we get the following evolution equations
\begin{equation}
3M^2_{Pl}H^2={}-\mathcal{M}^2\dot{\psi}\dot{\xi}+\Lambda-24H^3\dot{F}+\rho_m,
\label{einstein1}
\end{equation}
\begin{equation}
\label{einstein2}
\left(3H^2+2\dot{H}\right)M^2_{Pl}=\mathcal{M}^2\dot{\psi}\dot{\xi}+\Lambda-16H\left(H^2+\dot{H}\right)\dot{F}-8H^2\ddot{F}-P_m,
\end{equation}
where $F\equiv f(\psi(t))+\xi(t)$, and the dots denote the time derivatives.
The pressure $P_m$ and the energy density $\rho_m$ are components of the matter stress-energy tensor
\begin{equation}
\label{Tmunu}
{T_{m}}_{\mu\nu}={}-\frac{2}{\sqrt{-g}}\frac{\delta(\sqrt{-g}\mathcal{L}_m)}{\delta g^{\mu\nu}}
\end{equation}
that satisfy the energy conservation law
\begin{equation}
\dot{\rho}_m={}-3H(P_m+\rho_m).
\label{Equ_mat}
\end{equation}
In this paper, we consider only models with a perfect fluid, for which the EoS parameter $w_{\mathrm{m}}\equiv P_{\mathrm{m}}/\rho_{\mathrm{m}}$ is a constant. We introduce the cosmological constant $\Lambda$ as well. Thus, we can assume that $w_{\mathrm{m}}\neq -1$, without loss of generality.

Varying the  action $S_{L}$ over $\psi$, we get the following field equation:
\begin{equation}
\mathcal{M}^2\Box{\xi}=f'_{,\psi}\mathcal{G},
\label{equxi}
\end{equation}
where a prime denotes the derivative with respect to  $\psi$.

\section{Reconstruction procedure}
Our goal here is to find a function $f$ such that the model has a solution with a given behavior of the Hubble parameter.
If $H(t)$ is given, then   one can obtain $\psi(t)$, solving a linear differential equation~(\ref{equpsi}):
\begin{equation}
\mathcal{M}^2\left(\ddot\psi+3H\dot\psi\right)+24H^2\left(\dot H +H^2\right)=0.
\label{equpsim}
\end{equation}
If we know how $\rho_m$ depends on $P_m$, then, from Eq.~(\ref{Equ_mat}), we get $\rho_m(t)$ and $P_m(t)$.

Adding up Eqs.~(\ref{einstein1}) and (\ref{einstein2}), we obtain
a linear differential equation for~$F(t)$:
\begin{equation}
\label{equPsi} 8H^2\ddot{F}+8H\left(5H^2+2\dot H\right)\dot{F}=2\Lambda-2M^2_{Pl}\left(3H^2+\dot H\right)-P_{\mathrm{m}}+\rho_{\mathrm{m}}.
\end{equation}
Note that Eq.~(\ref{equPsi}) is the first order differential equation for $\dot{F}(t)$.

By solving this equation, we get the function $F(t)$. Further, substituting the relation $\xi(t)=F(t)-f(\psi(t))$ into Eq.~(\ref{equxi}) and using Eq.~(\ref{equpsim}), we obtain the following linear differential equation for $f(\psi)$:
\begin{equation}
\label{equf1}
   (\dot{\psi})^2 f''_{,\psi\psi}-48\frac{H^2}{\mathcal{M}^2}\left(\dot H +H^2\right)f'_{,\psi}=\ddot{F} +3H\dot{F}.
\end{equation}
To get $f(\psi)$ as a solution of Eq.~$(\ref{equf1})$ we invert $\psi(t)$ and obtain $t(\psi)$. For some important behaviors of the Hubble parameter this is possible to do.

\section{Search of de Sitter solutions}
Now, we look for a  function  $f(\psi)$ such that the model has de Sitter solutions.
So, $H$ is a nonzero constant: $H=H_0$.
In this case, the general solution of Eq.~(\ref{Equ_mat}) is
\begin{equation*}
    \rho_m =\rho_0 \mathrm{e}^{-3H_0(w_{\mathrm{m}}+1)t},
\end{equation*}
where $\rho_0$ is an integration constant.
Substituting the obtained function $\rho_m(t)$ and $H=H_0$ into Eq.~(\ref{equPsi}), we get the following solution
\begin{equation*}
    \dot{F}(t)=\left\{
    \begin{split}
    & \frac{(w_{\mathrm{m}}-1)\rho_0 \mathrm{e}^{-3H_0(w_{\mathrm{m}}+1)t}}{8(3w_{\mathrm{m}}-2)H_0^3}+\frac{(\Lambda-3M_{Pl}^2H_0^2)}{20H_0^3}+C_1\mathrm{e}^{-5H_0t},\quad
     w_{\mathrm{m}}\neq \frac{2}{3},\\
    &\frac{\rho_0 t \mathrm{e}^{-5H_0t}}{24H_0^2}+\frac{\Lambda-3M_{Pl}^2H_0^2}{20H_0^3}+\tilde{C}_1\mathrm{e}^{-5H_0t},\quad w_{\mathrm{m}}=\frac{2}{3},
    \end{split}
    \right.
\end{equation*}
where $C_1$ and $\tilde{C}_1$ are integration constants.

 Equation~(\ref{equpsim}) has
the following general solution:
\begin{equation*}
    \psi=\psi_1\mathrm{e}^{-3H_0(t-t_0)}-8\frac{H_0^3}{\mathcal{M}^2}(t-t_0),
\end{equation*}
where $\psi_1$ and $t_0$ are integration constants. Without loss of generality, we assume that $t_0=0$.
To get $t(\psi)$ in the explicit form, we set $\psi_1=0$ and obtain
    $$t={}-\frac{\mathcal{M}^2}{8H_0^3}\psi\,.$$

In the case of $w_{\mathrm{m}}\neq 2/3$, Eq.~\eqref{equf1}  has the following solution,
\begin{equation*}
    f(\psi)=\frac{2C_1}{5H_0}\,\mathrm{e}^{\frac {5\mathcal{M}^2}{8H_0^2}\psi}+
    f_1\mathrm{e}^{\frac{3\mathcal{M}^2}{4H_0^2}\psi}+\frac{\left(3M^2_{Pl}H_0^2-\Lambda\right)\mathcal{M}^2}{320H_0^6}\,\psi
    -\frac{\rho_0w_{\mathrm{m}}\,\mathrm{e}^{\frac{3(w_{\mathrm{m}}+1)\mathcal{M}^2}
    {8H_0^2}\psi}}{24H_0^4(3w_{\mathrm{m}}-2)(w_{\mathrm{m}}+1)},
\end{equation*}
where $f_1$ is an integration constant. The evolution equations define $f(\psi)$ up to a constant. So, we can put this constant equal to zero without loss of generality.

In the case of $w_{\mathrm{m}}= 2/3$, we obtain:
\begin{equation*}
    f(\psi)=\frac{2\tilde{C}_1}{5H_0}\, \mathrm{e}^{\frac {5\mathcal{M}^2}{8H_0^2}\psi}+
    f_1\,\mathrm{e}^{\frac{3\mathcal{M}^2}{4H_0^2}\psi}+\frac{\left(3M^2_{Pl}H_0^2-\Lambda\right)\mathcal{M}^2}{320H_0^6}\,\psi
    -\frac{5\mathcal{M}^2\psi+52H_0^2}{2400H_0^6}{
\,\rho_0 \mathrm{e}^{\frac {5\mathcal{M}^{2}}{8H_0^2}\psi}}.
\end{equation*}

Substituting the  functions here obtained into Eq.~(\ref{einstein1}), we get a restriction on the integration constant: $f_1=0$. This restriction is the same both for the case of $w=2/3$ and for the opposite case.

The obtained results motivate more detailed analysis of the considered nonlocal models with the function $f$, which is a sum of exponents. We leave the detailed study of the  de Sitter solutions obtained for future investigations.

\section*{Acknowledgements}

 E.E. is supported by MINECO (Spain), project FIS2016-76363-P, and AGAUR (Catalan Government), project 2017-SGR-247.
 E.O.P. and S.Yu.V. are supported in part by RFBR Grant No.~18-52-45016.

\end{document}